\documentclass[runningheads,fleqn]{llncs}

\usepackage[T1]{fontenc}
\def\doi#1{\href{https://doi.org/\detokenize{#1}}{\url{https://doi.org/\detokenize{#1}}}}

\usepackage{algorithm}
\usepackage[english]{babel}
\usepackage[noend]{algpseudocode}
\usepackage{graphicx}
\usepackage{listings}
\usepackage{xcolor}
\usepackage{amsmath}
\usepackage{sfmath} 
\usepackage{pgfplots}
\pgfplotsset{compat=1.17} 
\usepackage{booktabs} 
\usepackage{array}    
\usepackage{booktabs}
\usepackage{subcaption} 
\usetikzlibrary{pgfplots.statistics} 
\usepgfplotslibrary{groupplots} 
\pgfplotsset{compat=newest}
\usepackage[inkscapelatex=false]{svg}
\usepackage{tcolorbox}
\usepackage{listings}

\usepackage{tikz}
\usetikzlibrary{arrows.meta, positioning, shapes, fit}
\usepackage[hidelinks]{hyperref}
\usepackage{pgf-umlsd}
\usepackage{booktabs}
\usepackage{algorithm}
\usepackage[english]{babel}
\usepackage[noend]{algpseudocode}

\usepackage{tikz}
\usetikzlibrary{arrows.meta,positioning,fit,backgrounds,shapes.multipart,calc,decorations.pathmorphing,shapes.geometric,shapes.symbols}
\usepackage{pgfplots}
\usepackage{caption}
\usepackage{subcaption}
\usepackage{multirow}
\usepackage{color,soul}
\usepackage{enumitem}

\newcommand{\modelname}{APLiance}

\newif\ifMonochrome

\ifMonochrome
  \definecolor{C1}{rgb}{0,0,0}
  \definecolor{C2}{rgb}{0,0,0}
  \definecolor{C3}{rgb}{0,0,0}
  \definecolor{C4}{rgb}{0,0,0}
  \definecolor{C5}{rgb}{0,0,0}
  \definecolor{Bg}{gray}{0.93}
\else
  \definecolor{C1}{HTML}{1565C0} 
  \definecolor{C2}{HTML}{6A1B9A} 
  \definecolor{C3}{HTML}{2E7D32} 
  \definecolor{C4}{HTML}{EF6C00} 
  \definecolor{C5}{HTML}{AD1457} 
  \definecolor{Bg}{HTML}{F5F7FA} 
\fi

\newcommand{\att}[1]{\texttt{#1}}

\tikzset{
  >={Latex[length=2.5mm]},
  font=\sffamily,
  small/.style={font=\normalsize},
  note/.style={draw=none, align=center, inner sep=0pt, font=\normalsize},
  phase/.style={draw=C1!20, rounded corners=6pt, fill=Bg, inner sep=6pt},
  block/.style={draw=#1!60!black, fill=#1!6, thick, rounded corners=4pt,
                minimum width=32mm, minimum height=9mm, align=center},
  block/.default=C1,
  thinblock/.style={draw=#1!60!black, fill=#1!6, rounded corners=4pt,
                    minimum width=30mm, minimum height=7mm, align=center},
  io/.style={block=C1, minimum height=10mm},
  arrow/.style={C1, line width=0.9pt},
  dashedarrow/.style={C1, line width=0.9pt, dashed},
  titlebar/.style={draw=none, fill=white, align=left, inner sep=2pt, font=\bfseries\normalsize},
  lane/.style={rounded corners=4pt, draw=black!15, fill=black!2},
  kmeans/.style={draw, rounded corners=3pt, thick, fill=white, minimum height=32mm, minimum width=40mm},
  dataset/.style={draw, trapezium, trapezium angle=70, thick, fill=white, minimum height=10mm, minimum width=34mm},
  action/.style={draw, thick, rounded corners=3pt, fill=white, minimum height=10mm, minimum width=34mm},
  stage/.style={font=\bfseries\Large, align=center},
  line/.style={thick},
  faint/.style={draw=black!30},
  title/.style={font=\bfseries\large, text=black!60},
  neuron/.style={circle,minimum size=11pt,inner sep=0pt,draw=black!50,fill=black!15},
  in-neuron/.style={neuron, fill=green!45},
  hid-neuron/.style={neuron, fill=blue!45},
  out-neuron/.style={neuron, fill=red!55},
  vecbox/.style={draw, thick, minimum width=8mm, minimum height=5mm, fill=white, font=\small},
  vecdots/.style={draw=none, minimum width=4mm, minimum height=5mm, fill=none, font=\tiny}
}

\tikzset{
  kMeansDiagram/.pic={
    \begin{scope}[scale=0.8, transform shape, shift={(-1.5cm,-1.5cm)}]
      \foreach \x/\y in {0.5/1.2, 0.8/1.5, 1.2/1.1, 0.7/0.9, 1.0/1.3} {
        \fill[col1] (\x,\y) circle (2pt);
      }
      \node[star, star points=5, fill=col1, inner sep=2pt] at (0.9,1.2) {};
      \draw[dashed, col1] (0.9,1.2) circle (0.5);
    
      \foreach \x/\y in {2.0/2.3, 2.3/2.6, 2.7/2.2, 2.2/2.0, 2.5/2.5} {
        \fill[col2] (\x,\y) circle (2pt);
      }
      \node (-c2) [star, star points=5, fill=col2, inner sep=2pt] at (2.4,2.3) {};
      \draw[dashed, col2] (2.4,2.3) circle (0.5);
    
      \foreach \x/\y in {1.5/0.3, 1.8/0.5, 2.2/0.2, 1.7/0.1, 2.0/0.6} {
        \fill[col3] (\x,\y) circle (2pt);
      }
      \node[star, star points=5, fill=col3, inner sep=2pt] at (1.9,0.3) {};
      \draw[dashed, col3] (1.9,0.3) circle (0.5);
    
      \coordinate (-origin) at (0,0);
      \draw[->] (-origin) -- (0,3) node[left] {pc2};
      \draw[->] (-origin) -- (3,0) node (-pc1) [below] {pc1};
    \end{scope}
  }
}

\begin{document}
\title{Privacy as Permissible Operations: An ABAC Framework for Policy–Law Compliance}
\titlerunning{An ABAC Framework for Privacy Policy–Law Compliance}
\author{Ajay Dhakar\inst{1} 
Arunesh Sinha\inst{2} \and
Shamik Sural\inst{1} } 

\institute{Indian Institute of Technology Kharagpur, India\\
\email{ajaydhaker2002@gmail.com, shamik@cse.iitkgp.ac.in} \and
Rutgers University, USA\\
\email{arunesh.sinha@rutgers.edu}
}
\authorrunning{Dhakar
et al.}
\maketitle

\setcounter{footnote}{0}

\begin{abstract}
In recent years, many countries have started enacting laws to safeguard privacy of personal data of their citizens collected and maintained by various enterprises through  websites, mobile apps, and other means. It is imperative that the privacy policies of these enterprises respect the provisions of the applicable law. In this paper, we show how such organizational privacy policies can be efficiently checked against a prevalent law. Our novel approach named APLiance (\underline{A}BAC
framework for \underline{P}olicy–\underline{L}aw Compl\underline{iance}) models the requirements of the different sections of a privacy law in the form of  Attribute-based Access Control (ABAC) rules and the clauses of a privacy policy as a sequence of implied access requests. A policy is considered to be compliant with the law if these access requests are permitted by the corresponding ABAC rules. Although APLiance can be used in any policy-law setting, we demonstrate its effectiveness in the context of the recently introduced Digital Personal Data Protection Act of India. A browser plugin has been developed and publicly released for real time compliance checking using \modelname\ whenever a user visits the privacy policy page of a website.

\end{abstract}
\section{Introduction}
\label{sec:intro}
Over the last few years, increasing emphasis is being given to the privacy of online data shared by users with various enterprises. This is because users are often unable to fully grasp the quasi-legal language of the privacy policies presented to them and the full implication of accepting the same. Moreover, they might even be coerced into giving consent for data sharing as the services otherwise are not made available. To protect their citizens, different countries and states have enacted privacy laws that govern such personal data usage by enterprises. These include, among others, the European General Data Protection Regulation (GDPR)\footnote{\url{https://gdpr-info.eu/}}, Health Insurance Portability and Accountability Act (HIPAA)\footnote{\url{https://www.hhs.gov/hipaa/index.html}}, California Consumer Privacy Act (CCPA)\footnote{\url{https://oag.ca.gov/privacy/ccpa}}, and the recently introduced Digital Personal Data Protection (DPDP) Act\footnote{\url{https://www.meity.gov.in/static/uploads/2024/06/2bf1f0e9f04e6fb4f8fef35e82c42aa5.pdf}} of India. At the same time, there has been a significant rise in the number of online and mobile apps with which users interact on a daily basis. Most of them require new customers to first register with the app, share their personal details and accept a privacy policy, before any access to services or products is permitted. It is, therefore, expected that the conditions stated in a company's privacy policy that a user is supposed to accept, indeed comply with the extant privacy laws - a fact that should also be easily verifiable. 
In the rest of this paper, a \textit{privacy law} refers to the legal requirements (typically passed as an Act or Regulation) that are expected to be complied with in a given context, while a \textit{privacy policy} denotes the set of terms and conditions that an enterprise puts forth digitally for acceptance by a potential user of its products or services.  

Note that, privacy laws and organizational privacy policies ultimately govern one concrete question: Is a requested operation on personal data allowed in the present context? Indeed, past frameworks~\cite{barth2006privacy} have implicitly or explicitly reduced privacy reasoning to questions about the permissibility of information flows or data-access events under structured constraints. Whether expressed as logical rules over information flows, policy languages, or access-control conditions, these approaches share a common operational core: privacy requirements constrain which actions over data are admissible given contextual parameters. However, translating legal text and policy documents into a unified, machine-verifiable representation of such admissibility remains challenging. In particular, privacy laws are articulated in natural language with rich contextual qualifiers, while privacy policies describe intended data practices at varying levels of abstraction. 

In this paper, we propose a novel framework that attempts to close this gap by formalizing a privacy law as a set of Attribute-based Access Control (ABAC) rules \cite{abac}, and representing an organization's privacy policy as a collection of intended access requests. Compliance checking in the proposed framework, which we name as APLiance (\underline{A}BAC
framework for \underline{P}olicy–\underline{L}aw Compl\underline{iance}), thus reduces to a precise decision problem: Are the access requests specified in the policy permitted by the rules derived from the applicable law? It needs to be emphasized here that unlike access control systems, \textit{no access request is actually made in APLiance}. Rather, those are statically analyzed for any non-compliance. A closely related line of work is Contextual Integrity (CI)~\cite{nissenbaum2004privacy}, which models privacy in terms of the appropriateness of information flows under context-specific norms. \modelname\ can be viewed as an alternative formalization, capturing such norms within an ABAC framework to enable operational compliance checking. 

Although APLiance is a generic framework, we particularly focus on the DPDP Act of India. Such a choice is primarily motivated by the fact that while there is existing work on compliance checking for well-established acts like CCPA and GDPR, there is no work yet on DPDP.  Besides, with a significant majority of the most populous country in the world using scores of online apps daily, an approach towards mitigating the gap between current privacy practices and those demanded by the new law will have a major societal impact. 

While APLiance provides a precise formal representation for compliance checking, instantiating it in practice requires extracting attribute values corresponding to a given privacy policy. Since such policies are written in natural language and often exhibit variability in structure and terminology, we leverage Large Language Models (LLMs) to map policy text into the attribute space defined by our ABAC formulation. In particular, given a fixed attribute schema derived from the privacy law, the LLM is used to infer values of base attributes from the policy text, including identifying cases where attribute values are not explicitly specified and must be treated as unknown. This separation between formal rule specification and LLM-based attribute extraction allows us to combine the robustness and interpretability of rule-based compliance checking with the flexibility of modern language models in processing unstructured policy documents. To further establish the practicality of our work, we have built a browser plugin that can in real time check for compliance of a policy and the various clauses of the DPDP act, whenever a user visits an enterprise's privacy policy web page.
In summary, this paper makes the following novel contributions.

\begin{enumerate}[label=\roman*]
    \item APLiance is the first ever work on DPDP Act compliance checking.
    \item Both privacy laws and privacy policies are modeled in an ABAC framework that makes analysis easier and explainability more transparent.
    \item Effective use of LLMs for checking the validity of a privacy policy with respect to a privacy law. 
    \item Development of a browser extension for real time compliance checking.
\end{enumerate}

The rest of the paper is structured as follows. In the next section, we discuss some of the preliminary concepts related to ABAC and the DPDP Act. The architecture of \modelname\ is introduced in Section \ref{sec:arch} followed by its design and implementation details in Section \ref{sec:system_overview}. We present results of our experiments in Section \ref{sec:results}. Related work is reviewed in Section \ref{sec:related_work} and we conclude in Section \ref{sec:conclusion} with some directions for future work. 

\section {Preliminaries}
\label{sec:prelim}

In this section, we introduce some of the basic concepts of ABAC and present a short description of the DPDP Act.
\subsection{Attribute-based Access Control}
\label{subsec:prelimabac}
ABAC is an access control paradigm in which authorization decisions are determined by evaluating attributes associated with entities involved in a request, rather than relying on fixed identities or predefined roles.
Let $\mathcal{S}$, $\mathcal{O}$, $\mathcal{A}$ and $\mathcal{E}$ denote the sets of subjects, objects, actions and environmental contexts, respectively. An access request is defined as a tuple
\[
t = (s, o, a, e) \in \mathcal{S} \times \mathcal{O} \times \mathcal{A} \times \mathcal{E}.
\]

Each component of a request $t$ is associated with a collection of attribute values describing its relevant properties. The union of these values constitutes the attribute description of the request, which we denote by $\mathbf{v}$.
An ABAC policy is a function
\[
\pi : \mathcal{S} \times \mathcal{O} \times \mathcal{A} \times \mathcal{E} \to \{\texttt{permit}, \texttt{deny}\},
 \text{ where }\]
\[
\pi(t) =
\begin{cases}
\texttt{permit} & \text{if } \varphi(\mathbf{v}, a) \text{ holds},\\
\texttt{deny} & \text{otherwise},
\end{cases}
\]
and $\varphi$ is a predicate (rules) over the attribute description $\mathbf{v}$ and the action $a$.
For example, a rule $r$ in an ABAC policy may specify that a request is permitted if
\[
\texttt{designation} = \texttt{doctor}
\;\wedge\;
\texttt{type} = \texttt{medical\_record},
\]
and denied otherwise. Given a request $t$, the authorization decision is obtained by evaluating $\pi(t)$. 

ABAC enables fine-grained and context-dependent access control by reasoning directly over attributes, making it well-suited for dynamic environments and for settings in which policies must encode regulatory or organizational constraints. We utilize this formalism as an operational model for determining if a given data access request is permitted under a specified law.

\subsection{Digital Personal Data Protection Act of India}
\label{subsec:prelimdpdp}

The DPDP Act passed in the year 2023 is India's first comprehensive law dedicated to digital privacy. It provides a framework to regulate the processing of digital personal data, mentioning two key objectives: protecting an individual's right to privacy, and enabling lawful data processing by organizations. The Act establishes a comprehensive legal framework governing how personal data is collected, used, stored, and shared. 

The DPDP Act lays down principles to ensure transparency, trust and responsible data handling practices. It defines obligations for entities (data fiduciaries) processing personal data, including requirements for obtaining valid consent, specifying clear purpose for data usage, enabling withdrawal of consent, and ensuring grievance redressal mechanisms. While the act spans multiple chapters, provisions relevant to privacy policies are primarily contained up to Chapter~IV (till Section~17), which focuses on consent, rights of individuals, and lawful processing conditions. In this work, however, we formalize rules only up to Section~7, as these capture the most critical aspects of consent and data processing conditions for compliance checking. This choice aligns with our primary goal of developing an extensible modeling framework, rather than exhaustively encoding all provisions of the Act.

\section{Overall Framework of \modelname}
\label{sec:arch}

In \modelname, we model privacy constraints as governing the admissibility of data operations under context. Expanding on the notations introduced in Sub-section \ref{subsec:prelimabac}, let $\mathcal{R}$ denote a finite set of roles (equivalently, these constitute a set of possible values for a subject attribute). For each role $r \in \mathcal{R}$, let $\mathcal{S}_r$ denote the corresponding set of subjects.
A request is defined as a tuple
\[
t = \big( (s_r)_{r \in \mathcal{R}},\ o,\ a,\ e \big) \in \left( \prod_{r \in \mathcal{R}} \mathcal{S}_r \right) \times \mathcal{O} \times \mathcal{A} \times \mathcal{E},
\]
where $s_r \in \mathcal{S}_r$ denotes the entity playing role $r$, $o \in \mathcal{O}$ is the data object, $a \in \mathcal{A}$ is the action, and $e \in \mathcal{E}$ represents the environmental context.

Associated with each request, is a collection of attribute values describing the relevant properties of its components (subjects in roles, data object, and environment). We denote this attribute description by $\mathbf{v}$, and let $\mathcal{V}$ denote the space of all such attribute assignments. 
We distinguish three types of attributes:
\begin{itemize}[label=\textbullet]
    \item \emph{Base attributes} - values of these attributes are directly extracted from a privacy policy or system description
    \item \emph{Derived attributes} - values of these attributes are determined from other attributes using explicit rules
    \item \emph{Unknown attributes} - values of these attributes are not specified and may take any value in their domain.
\end{itemize}
A \emph{derivation rule} has the form: 
\[
\text{if condition } C \text{ holds, then attribute } \alpha \text{ takes value } c,
\]
where $C$ is a condition over attribute values. These rules are used to compute derived attributes from base attributes. For example, a rule may state that if all required items are present in a consent notice, then the attribute \texttt{consent\_notice\_wellformed} is set to \texttt{true}.

Starting from the base attributes associated with a request, we repeatedly apply all relevant derivation rules until no further attribute values can be determined. This produces an enriched attribute description of the request, consisting of the original base attributes together with all derived attributes implied by the rules. Such a use of derived attributes enables modular specification, allowing complex conditions to be encapsulated and reused across multiple rules without duplicating their underlying structure.

A \emph{decision rule} has the form:
\[
\text{if condition } D \text{ holds, then action } a \text{ is permitted}
\]
where $D$ is a condition over the attribute description $\mathbf{v}$. The collection of all such rules determines the policy decision for the request.
If some attributes are unknown, we evaluate the request under all possible values of those attributes that are consistent with their domains. A request is called \emph{admissible} if it is permitted in every such case. It is called \emph{violating} if there exists at least one such case in which it is denied.

Next, we present examples based on the DPDP Act. In these examples, we consider an attribute called $role$ that takes values such as (data) \texttt{fiduciary}, corresponding to the entity responsible for data processing decisions, and (data) \texttt{principal}, corresponding to the individual whose data is being processed. Requests involve these roles together with actions such as \texttt{data\_processing}, and are evaluated based on the corresponding attribute description $\mathbf{v}$.
\paragraph{Example 1 (Consent-based processing).}
Let $\mathcal{R} = \{\texttt{fiduciary}, \texttt{principal}\}$ and consider the action $a = \texttt{data\_processing}$. Suppose the relevant attributes include $\texttt{consent\_status}$ (associated with the principal) and $\texttt{lawful\_purpose}$ (environmental). A decision rule may specify that processing is permitted if
\[
\texttt{consent\_status} = \texttt{active} \;\wedge\; \texttt{lawful\_purpose} = \texttt{true}.
\]

\paragraph{Example 2 (Derived attributes and modularity).}
Suppose $\texttt{consent\_preconditio}\\$$\texttt{ns\_fulfilled}$ is a derived attribute defined by rules requiring multiple conditions, such as the presence of a well-formed notice, informed consent, and the availability of a withdrawal mechanism. Instead of repeating all these conditions in every decision rule, one may use the derived attribute and specify that processing is permitted only if $\texttt{consent\_preconditions\_fullfilled} = \texttt{true}$. The underlying derivation rules ensure that this attribute is set correctly based on the base attributes, thereby enabling modular and reusable policy specifications.


\paragraph{Example 3 (Unknown attributes).}
Consider the attribute $\texttt{reasonable\_time\_ela}$\\$\texttt{psed}$, which depends on external temporal information and may not be specified in the privacy policy. Suppose we have a decision rule that permits processing if
\[
(\texttt{consent\_status} = \texttt{withdrawn} \;\wedge\; \texttt{reasonable\_time\_elapsed} = \texttt{false})
\;\;\vee\;\;\]
\[\texttt{legitimate\_use} = \texttt{true}.
\]
If $\texttt{reasonable\_time\_elapsed}$ is unknown, we evaluate the rule under both the possible values $\{\texttt{true}, \texttt{false}\}$. If the rule yields different outcomes across these possibilities (e.g., permitted when $\texttt{reasonable\_time\_elapsed} = \texttt{false}$ but denied when it is $\texttt{true}$), then the request is not admissible. This reflects the need to account for all possible interpretations of attributes whose values are not determined by a privacy policy.

\section{Design Details and Implementation}
\label{sec:system_overview}

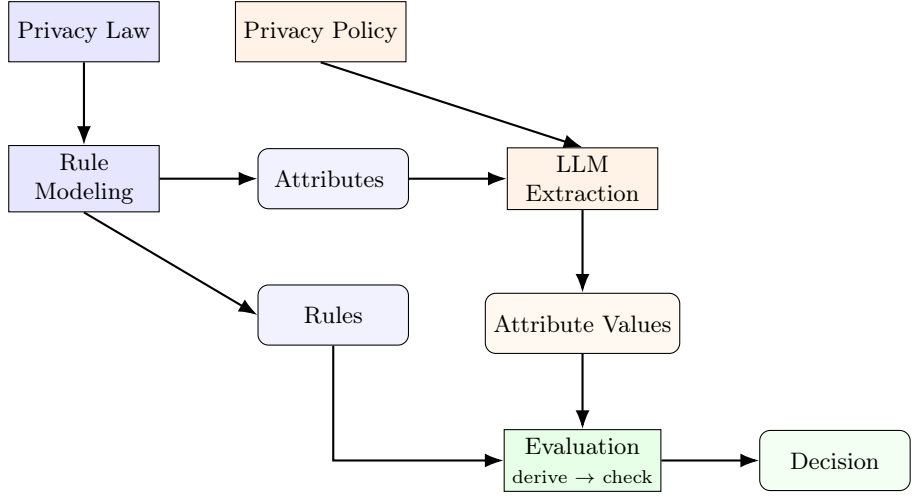
\begin{figure}[t]
\centering
\begin{tikzpicture}[
    node distance=1.1cm and 1.3cm,
    every node/.style={font=\footnotesize},
    box/.style={draw, rounded corners, align=center, minimum width=2.0cm, minimum height=0.8cm},
    proc/.style={draw, align=center, minimum width=2.0cm, minimum height=0.8cm},
    arrow/.style={->, thick}
]

\node[proc, fill=blue!10] (law) {Privacy Law};
\node[proc, right=1.0cm of law, fill=orange!10] (policy) {Privacy Policy};

\node[proc, below=of law, fill=blue!10] (model) {Rule\\Modeling};

\node[box, right=of model, fill=blue!5] (attr) {
Attributes
};

\node[proc, right=of attr, fill=orange!10] (llm) {LLM\\Extraction};

\node[box, below=1.0cm of attr, fill=blue!5] (rules) {Rules};
\node[box, below=of llm, fill=orange!5] (values) {Attribute Values};

\node[proc, below=1.0cm of values, fill=green!10] (eval) {
Evaluation\\{\scriptsize derive $\rightarrow$ check}
};

\node[box, right=of eval, fill=green!5] (decision) {Decision};

\draw[arrow] (law.south) -- (model.north);

\draw[arrow] (model.east) -- (attr.west);
\draw[arrow] (model.south) -- (rules.west);
\draw[arrow] (attr.east) -- (llm.west);

\draw[arrow] (policy.south) -- (llm.north);

\draw[arrow] (llm.south) -- (values.north);

\draw[arrow] (values.south) -- (eval.north);
\draw[arrow] (rules.south) |- (eval.west);

\draw[arrow] (eval.east) -- (decision.west);

\end{tikzpicture}

\caption{System architecture: legal rules are formalized into attribute-based constraints, policy text is mapped to attribute values using an LLM guided by the attribute schema, and compliance is determined via rule-based evaluation.}
\label{fig:sysarch}
\end{figure}

The \modelname\ 
architecture provides an end-to-end pipeline that checks to see if privacy policies follow the DPDP Act. As shown in Figure \ref{fig:sysarch}, the first step is a one-time manual inspection of the act that extracts attributes and rules from the privacy law, after defining the subjects. 
In such extraction, some attributes are labeled as \textit{base attributes} that can be directly inferred from privacy policy text. Also, some attributes are labeled as \textit{unknown attributes} that stand for differing scenarios that can happen in the future. The manual extraction step also labels some attributes as \textit{derived} and such labeling naturally follows from the structure of the DPDP act itself\footnote{\url{https://www.dpdpa.com/}}.
Next we use an LLM to look at the privacy policy text of the target website. The LLM is prompted to figure out \textit{base attributes} directly from the policy text. The rule engine checks these attributes against codified DPDP rules to determine the final compliance verdict. Lastly, the system makes violation reports and explanations that can be acted upon. We have made a browser extension that shows these reports to end users so that they can easily check if the websites they visit are compliant with the law.


\subsection{\modelname\ Specification Extracted from Privacy Law}
\label{subsec:normal_attributes}
As mentioned before, the DPDP Act is comprised of multiple sections addressing different aspects of data regulation. However, not all provisions are directly relevant to privacy policy compliance. A privacy policy is a user-facing document that communicates key elements such as consent, purpose of data collection, and data usage practices.
Other provisions of the Act impose organizational obligations, such as data security and internal compliance measures, which do not necessarily need to be explicitly included in a privacy policy. Therefore, this analysis focuses on Sections 1–7 of the DPDP Act, as these are most relevant to user consent and data rights, and thus directly impact privacy policy requirements.

\paragraph{Metadata} We extract two values for the attribute \textit{role}, namely, \att{data\_fiduciary} and \att{data\_principal}. The object is the data owned by the subject in role  \att{data\_principal} and the action \att{data\_processing} is requested by the subject in role \att{data\_fiduciary}.

\paragraph{Base Attributes} Base attributes are termed \textit{base} because they correspond to concrete, observable textual facts or omissions within the document itself. 
When explaining each attribute, we focus on its legal significance, how it typically manifests in a privacy policy text, and whether an omission should be treated as \textit{false} or \textit{unknown}. Base attributes in our framework are described below categorized into the the entity these are the attributes of.

\noindent \textbf{Attributes of subject with roles \att{data\_fiduciary}}:
\begin{itemize}[label=\textbullet]
    \item \att{offering\_service\_to\_data\_principal\_within\_india:} Checks if offering service to consumers or data principals based in India.

    \item \att{consent\_is\_freely\_given:} Whether the consent is freely given.
    \item \att{consent\_is\_specific\_to\_purpose:} Whether the consent is limited to the processing of personal data which is necessary for such a specified purpose
    \item \att{consent\_is\_informed:} Whether the data principal is informed of the Nature of personal data collected, Purpose of processing, Rights available (access, correction, withdrawal) and Contact details of grievance redressal.
    \item \att{consent\_is\_unambiguous:} Whether consent is expressed through a clear affirmative act.
    \item \att{consent\_request\_language\_all\_eighth\_schedule:} Whether the consent request is provided in English and in all languages listed in the Eighth Schedule of the Indian Constitution.
    \item \att{consent\_request\_contains\_contact\_details\_of\_dpo\_or\_equivalent:}\\ Whether the contact information of a data protection officer or person, who is able to answer on behalf of the data fiduciary about the processing of their personal data, is published.
    \item \att{easy\_consent\_withdrawal:} Whether the mechanism to withdraw consent is as easy and simple as the mechanism used to give consent.

    \item \att{consent\_notice\_information\_about\_personal\_data:} If the consent notice describes what personal data will be collected.
    \item \att{consent\_notice\_purpose\_of\_processing:} If the consent notice describes the purpose of collecting personal data.
    \item \att{consent\_notice\_how\_to\_exercise\_rights\_sec6.4:} If the consent notice describes how consent can be withdrawn.
    \item \att{consent\_notice\_how\_to\_exercise\_rights\_sec13:} If the consent notice describes means of grievance redressal provided by a Data Fiduciary in respect of any act or omission related to this law.
    \item \att{consent\_notice\_how\_to\_complaint\_to\_board:} If the consent notice describes how to complain to the Board.
\end{itemize}



\noindent \textbf{Attributes of object (data)}:
\begin{itemize}[label=\textbullet]
    \item \att{purpose\_of\_processing:} Purpose for which data is being processed/shared by data fiduciary.
\end{itemize}

\noindent \textbf{Attributes of environment}:
\begin{itemize}[label=\textbullet]
    \item \att{lawful\_purpose:} Any purpose which is not expressly forbidden by law.
\end{itemize}
\paragraph{Derived Attributes}
Unlike base attributes, these are not directly inferred from the policy text; instead they are inferred logically through derivation rule evaluation. 

\begin{itemize}[label=\textbullet]
    \item \att{law\_applicable:} Indicates whether the DPDP Act applies to the Data processing activity or not.
    \item \att{consent\_status:} Represents the current state of user consent, categorized as not\_given/active/withdrawn based on user actions and preconditions.
    \item \att{legitimate\_use:} Indicates whether organizations are allowed to process personal data without explicit consent in specific situations.
    \item \att{consent\_notice\_wellformed:} Indicates whether consent notice satisfies all requirements under DPDP act.
    \item \att{option\_for\_consent\_withdrawal:} specifies whether valid and accessible mechanism exists for data principal to withdraw consent.
    \item \att{past\_processing\_with\_active\_consent\_legal:} indicates whether data processing performed prior to consent withdrawal remains lawful.
    \item \att{consent\_preconditions\_fulfilled:} Indicates whether all conditions for a valid consent have been satisfied.
    \item \att{allow\_data\_processing:} Final attribute, indicates whether the data processing is legally permissible under the evaluated conditions.
\end{itemize}

\paragraph{Unknown Attributes}
These are defined as runtime facts that cannot be inferred from the privacy policy directly.
\begin{itemize}[label=\textbullet]
    \item \att{voluntary\_data\_for\_specified\_purpose:} Whether the user has voluntarily provided personal data for the defined purpose.
    \item \att{consent\_action:} Represents whether the user has explicitly provided consent for data processing.
    \item \att{consent\_withdraw\_action:} Represents whether the user has explicitly indicated her intent to withdraw previously given consent.
    \item \att{reasonable\_time\_elapsed:} Indicates whether the reasonable time has passed since consent withdrawal, after which processing is no longer permitted.
    \item \att{consent\_for\_state\_benefits:} Whether user has provided consent for processing related to state benefits or services.
    \item \att{available\_to\_state\_and\_notified\_by\_government:} Indicates whether the personal data is already available with the state and has been notified by the government.
\end{itemize}

\noindent\textit{Rules Extracted from Privacy Law} We next list out the rules extracted from the DPDP Act.\\

\noindent\textbf{Rules 1, 2 ($R_{1}$, $R_{2}$) (Law applicability)}
\setlength{\mathindent}{0pt}
{ \small
\begin{equation*}
\begin{array}{l l}
\mathbf{if}   \quad  & \att{offering\_service\_to\_data\_principal\_within\_india}
                       = \att{true}  \\
\mathbf{then} \quad & \att{law\_applicable} \leftarrow \att{true} 
\end{array}
\end{equation*}
}
{ \small
\begin{equation*}
\begin{array}{l l}
\mathbf{if}   \quad & \att{offering\_service\_to\_data\_principal\_within\_india}
                       = \att{false} \\
\mathbf{then} \quad & \att{law\_applicable} \leftarrow \att{false}
\end{array}
\end{equation*}
}%
\noindent\textbf{Rule 3 ($R_{3}$) (Processing permission)}
{ \small
\begin{equation*}
\begin{array}{l l}
\mathbf{if}   \quad
  & \att{law\_applicable} = \att{true} \;\wedge\; \att{lawful\_purpose} = \att{true} \\
  & \;\wedge\;
    \big(\,\att{consent\_status} = \att{active}
    \;\vee\;
    \att{legitimate\_use} = \att{true}\,\big) \\
\mathbf{then} \quad
  & \att{allow\_data\_processing} \leftarrow \att{true}
\end{array}
\end{equation*}
}%
\noindent\textbf{Rule 4 ($R_{4}$) (Consent notice completeness)}
{ \small
\begin{equation*}
\begin{array}{l l}
\mathbf{if}   \quad
  & \att{consent\_notice\_information\_about\_personal\_data} = \att{true} \\
  & \;\wedge\; \att{consent\_notice\_purpose\_of\_processing} = \att{true} \\
  & \;\wedge\; \att{consent\_notice\_how\_to\_exercise\_rights\_sec6.4}
               = \att{true} \\
  & \;\wedge\; \att{consent\_notice\_how\_to\_exercise\_rights\_sec13}
               = \att{true} \\
  & \;\wedge\; \att{consent\_notice\_how\_to\_complaint\_to\_board}
               = \att{true} \\
  & \;\wedge\; \att{notice\_languages\_all\_eighth\_schedule}
               = \att{true} \\
\mathbf{then} \quad
  & \att{consent\_notice\_wellformed} \leftarrow \att{true}
\end{array}
\end{equation*}
}%
\noindent\textbf{Rule 5 ($R_{5}$) (Valid consent)}
{ \small
\begin{equation*}
\begin{array}{l l}
\mathbf{if}   \quad
  & \att{consent\_is\_freely\_given} = \att{true} \\
  & \;\wedge\; \att{consent\_is\_specific\_to\_purpose} = \att{true} \\
  & \;\wedge\; \att{consent\_is\_informed} = \att{true} \;\wedge\; \att{consent\_is\_unambiguous} = \att{true} \\
  & \;\wedge\; \att{consent\_request\_language\_all\_eighth\_schedule}
               = \att{true} \\
  & \;\wedge\; \att{consent\_request\_contains\_contact\_details\_of\_dpo\_or\_equivalent}
               = \att{true} \\
  & \;\wedge\; \att{option\_for\_consent\_withdrawal} = \att{true} \\
  & \;\wedge\; \att{consent\_notice\_wellformed} = \att{true} \\
\mathbf{then} \quad
  & \att{consent\_preconditions\_fullfilled} \leftarrow \att{true}
\end{array}
\end{equation*}
}%
\noindent\textbf{Rule 6 ($R_{6}$) (Withdrawal option)}
{ \small
\begin{equation*}
\begin{array}{l l}
\mathbf{if}   \quad 
& \att{easy\_consent\_withdrawal} = \att{true} \\
\mathbf{then} \quad & \att{option\_for\_consent\_withdrawal}
                       \leftarrow \att{true}
\end{array}
\end{equation*}
}%
\noindent\textbf{Rules 7, 8 ($R_{7}$, $R_{8}$) (Consent status)}
{ \small
\begin{equation*}
\begin{array}{l l}
\mathbf{if}   \quad
  & \att{consent\_preconditions\_fullfilled} = \att{true} \\
  & \;\wedge\; \att{consent\_action} = \att{true} \;\wedge\; \att{consent\_withdraw\_action} = \att{false} \\
\mathbf{then} \quad
  & \att{consent\_status} \leftarrow \att{active}
\end{array}
\end{equation*}
}%
{ \small
\begin{equation*}
\begin{array}{l l}
\mathbf{if}   \quad
  & \att{consent\_preconditions\_fullfilled} = \att{true} \\
  & \;\wedge\; \att{consent\_action} = \att{true} \;\wedge\; \att{consent\_withdraw\_action} = \att{true} \\
\mathbf{then} \quad
  & \att{consent\_status} \leftarrow \att{withdrawn}
\end{array}
\end{equation*}
}%
\noindent\textbf{Rule 9 ($R_{9}$) (Non-retroactivity)}
{ \small
\begin{equation*}
\begin{array}{l l}
\mathbf{if}   \quad & \att{consent\_status} = \att{withdrawn} \\
\mathbf{then} \quad & \att{past\_processing\_with\_active\_consent\_legal}
                       \leftarrow \att{true}
\end{array}
\end{equation*}
}%
\noindent\textbf{Rule 10 ($R_{10}$) (Post-withdrawal processing)}
{ \small
\begin{equation*}
\begin{array}{l l}
\mathbf{if}   \quad
  & \att{law\_applicable} = \att{true}  \;\wedge\; \att{lawful\_purpose} = \att{true} \\
  & \;\wedge\;
    \Big(\,\big(\,\att{consent\_status} = \att{withdrawn}
    \;\wedge\;
    \att{reasonable\_time\_elapsed} = \att{false}\,\big) \\
  & \qquad\;\vee\;
    \att{legitimate\_use} = \att{true}\Big) \\
\mathbf{then} \quad
  & \att{allow\_data\_processing} \leftarrow \att{true}
\end{array}
\end{equation*}
}%
\noindent\textbf{Rule 11 ($R_{11}$) (Legitimate use)}
{ \small
\begin{equation*}
\begin{array}{l l}
\mathbf{if} \quad
  & \big(\,\att{consent\_status}
           \in \{\att{not\_given},\;\att{active}\} \\
  & \qquad \wedge\;
    \att{voluntary\_data\_for\_specified\_purpose} = \att{true}\big) \\[2pt]
  & \;\vee\;
    \big(\,(\,\att{consent\_for\_state\_benefits} = \att{true} \\
  & \qquad\quad
    \vee\; \att{available\_to\_state\_and\_notified\_by\_government}
           = \att{true}\,) \\
  & \qquad \wedge\;
    \att{purpose\_of\_processing} = \att{state\_benefits}\big) \\[2pt]
  & \;\vee\;
    \att{purpose\_of\_processing} \in
    \big\{\,\att{integrity\_of\_india},\;
            \att{sovereignty\_of\_india},\\
  & \qquad
    \att{security\_of\_state},\;
    \att{obligation\_under\_law},\;
    \att{medical\_emergency},\\
  & \qquad
    \att{disaster\_management},\;
    \att{safeguarding\_employment}\,\big\} \\[4pt]
\mathbf{then} \quad
  & \att{legitimate\_use} \leftarrow \att{true}
\end{array}
\end{equation*}
}

The above set of rules forms the basis for compliance checking with the published privacy policies of the individual enterprises offering services or products.
\subsection{Browser Extension}
\label{sec:browser_extension}

To demonstrate the practical applicability of \modelname, we developed a browser extension\footnote{\url{https://github.com/ajaydhaked/privacy-policy-compliance-extension}} that enables end users to assess the compliance of privacy policies with the DPDP Act in real time. The extension interacts with a remote backend\footnote{\url{https://github.com/ajaydhaked/privacy-policy-compliance-model-backend}} that encapsulates the LLM-based inference and rule evaluation pipeline. Compliance check can be initiated by clicking on the extension icon and selecting the \textit{Check Compliance} option.
Upon invocation, the extension extracts the textual content of the active webpage and sends the policy text, along with the page URL and title, to the backend via an HTTP POST request. The backend is deployed on \textit{PythonAnywhere}, and exposes an \texttt{/analyze} endpoint to handle compliance requests.
To improve efficiency and reduce redundant LLM API calls, the backend maintains a persistent disk cache indexed by the URL, page title, and a hash of the page content. Cached entries are assigned a time-to-live (TTL) of 24 hours.
The backend processes the request and returns a compliance verdict along with a set of identified violations. Each violation corresponds to an unsatisfied DPDP compliance attribute.
Upon receiving the response, the extension presents a clear verdict, COMPLIANT or NON-COMPLIANT, along with a numbered list of specific violations, enabling users to easily understand the compliance gaps in the privacy policy.

\section{Experimental Results}
\label{sec:results}
In this section, we first introduce our experimental setup including the datasets, and then present our detailed results. Note that, in order that the results of our analysis do not construe any form of perceived maligning, we have anonymized the names of the companies whose privacy policies were used for testing. 

\subsection{Experimental Setup}

We construct a dataset of $25$ real-world privacy policies by collecting them from the official websites of organizations across domains such as e-commerce, social media, news and streaming platforms. Each policy is evaluated against a set of compliance attributes derived from the DPDP Act.
Ground truth annotations are constructed for $25 \times 16 = 400$ (policy, attribute) pairs, 
where each pair represents whether a specific compliance attribute 
(e.g., \texttt{consent\_is\_informed}) is satisfied by a given privacy policy. 
Each pair is labeled with one of $\{\texttt{true}, \texttt{false},\\ \texttt{unknown}\}$. 
The system predicts the same label space, and performance is evaluated using standard metrics over these instances.
For attribute extraction from privacy policy text, we use the \texttt{gpt-5.4-mini} model. The model is prompted with the predefined base attributes schema and is tasked with inferring values for base attributes from policy text (Details in  Appendix). All experiments are conducted using this model configuration without task-specific fine-tuning.

\subsection{Attribute Classification Performance}

\begin{table}[t]
\centering
\caption{Overall performance of \textsc{APLiance}}
\begin{tabular}{lcccc}
\toprule
Total & Accuracy & Precision & Recall \\
\midrule
400 & 0.95 & 0.96 & 0.96 \\
\bottomrule
\end{tabular}
\label{tab:overall}
\end{table}

\noindent
As shown in Table \ref{tab:overall}, \textsc{APLiance} achieves an accuracy of $94.5\%$ in base attribute-level  classification, with both precision and recall equal to $96.1\%$. These metrics are computed over the $400$ instances mentioned above. The confusion matrix consists of $272$ true positives, $106$ true negatives, $11$ false positives, and $11$ false negatives, indicating that the system makes a comparable number of false positive and false negative errors, maintaining a low mis-classification rate of $5.5\%$.

\begin{table*}[t]
\centering
\caption{Per-policy performance (Detailed results in Appendix Table \ref{tab:compoldetail})}
\begin{tabular}{lcccc}
\toprule
Policy & Accuracy & Precision & Recall \\
\midrule
A1 & 0.94 & 1.00 & 0.92 \\
F1 & 1.00 & 1.00 & 1.00 \\
G1 & 0.94 & 1.00 & 0.92 \\
L1 & 0.88 & 0.92 & 0.92 \\
N1 & 0.88 & 0.85 & 1.00 \\
W1 & 0.88 & 1.00 & 0.83 \\
Z1 & 0.81 & 0.75 & 1.00 \\
\bottomrule
\end{tabular}
\label{tab:perpolicyanalysis}
\end{table*}
\noindent
\paragraph{Per-Policy Analysis}
Next, we perform per-policy analysis and the results are shown in Table \ref{tab:perpolicyanalysis}. It is observed that \textsc{APLiance} demonstrates consistently strong performance across diverse real-world privacy policies. For each policy, we compute accuracy as the proportion of DPDP base attributes whose predicted labels match the ground truth annotations. Most policies achieve a per-policy base attribute classification accuracy above $93\%$, and several policies are classified correctly for all evaluated base attributes. Lower accuracy/recall for certain policies (W1 AND Z1) indicates a higher rate of mis-classification of individual compliance attributes, often due to ambiguous or implicit policy language.

\paragraph{Attribute-Level Analysis}
\begin{table*}[t]
\centering
\caption{Performance per base attribute (Detailed results in Appendix Table \ref{tab:perattribdetail})}
\begin{tabular}{lccc}
\toprule
Attribute & Accuracy & Precision & Recall \\
\midrule
consent\_is\_informed & 1.00 & 1.00 & 1.00 \\
consent\_is\_specific\_to\_purpose & 1.00 & 1.00 & 1.00 \\
lawful\_purpose & 1.00 & 1.00 & 1.00 \\
purpose\_of\_processing & 1.00 & 1.00 & 1.00 \\
consent\_is\_freely\_given & 0.96 & 0.96 & 1.00 \\
consent\_notice\_rights\_sec13 & 0.92 & 1.00 & 0.91 \\
consent\_notice\_rights\_sec6.4 & 0.88 & 0.88 & 1.00 \\
easy\_consent\_withdrawal & 0.76 & 0.57 & 0.57 \\
consent\_is\_unambiguous & 0.64 & 0.25 & 0.14 \\
\bottomrule
\end{tabular}
\label{tab:perattribanalysis}
\end{table*}
\noindent
Attribute-level analysis as shown in Table \ref{tab:perattribanalysis} reveals that explicitly stated requirements such as purpose limitation and lawful purpose achieve perfect performance. In contrast, attributes requiring interpretation of implicit language, such as whether consent is unambiguous or easily withdrawable, remain challenging.

\subsection{Privacy Policy Compliance}

Our evaluation as presented in the last sub-section shows that none of the analyzed privacy policies satisfies the full set of ABAC rules required to authorize lawful data processing under the DPDP Act. In particular, for every policy, at least one prerequisite condition, such as valid consent, legitimate use, or lawful purpose, was not met, preventing the inference of the attribute \att{allow\_data\_processing} as \att{true}. This outcome is consistent with the fact that enactment of the DPDP Act is quite recent, and organizations are still in the process of aligning their privacy practices with the new regulatory requirements.

A key observation is that, under the DPDP formulation, compliance is sensitive to violations of base attributes. Specifically, since each decision rule is an implication whose antecedent is a conjunction over base and dependent attributes, the presence of any violated prerequisite falsifies the antecedent and is sufficient to prevent compliance.
To better understand the sources of non-compliance, we analyze the impact of individual attributes.

The results show that no policy in our dataset is compliant, even with respect to the ground-truth annotations. This is primarily due to three attributes: \att{consent\_request\_language\_all\_eighth\_schedule},  
\att{notice\_languages\_all\_\\eighth\_schedule} and \att{consent\_notice\_how\_to\_complaint\_to\_board}. 

The first two attributes are false for all policies, as none of the analyzed privacy policies provides notices in all languages listed in the Eighth Schedule of the Constitution of India. The third attribute is only satisfied by 2 out of 25 policies, as most organizations do not yet specify mechanisms for filing complaints with the Data Protection Board. These observations are expected, given that the DPDP Act was enacted only recently and compliance requirements are still being operationalized in practice.

To isolate the effect of these attributes, we next bypassed them by excluding from the evaluation. Even under this relaxed setting, only 3 out of 25 policies are found to be compliant. The primary sources of non-compliance in this case are the attributes \att{consent\_is\_unambiguous} and \att{easy\_consent\_withdrawal}. In practice, many policies rely on implicit consent mechanisms (e.g., assuming consent through use of the service), which violates the requirement of clear affirmative action. Similarly, withdrawal of consent is often significantly more complex than providing consent, typically requiring manual processes such as contacting a data protection officer, therefore failing the DPDP Act requirement that withdrawal be as easy as giving consent. This outcome also reflects the inherent ambiguity in policy language, particularly for attributes such as \att{consent\_is\_unambiguous} and \att{easy\_consent\_withdrawal}, which are often difficult to assess consistently even for human readers.

Finally, when these two additional attributes are also relaxed, we observe that 20 out of 25 policies satisfy the remaining conditions. This suggests that a large fraction of organizations already meet several structural requirements of the DPDP Act. Overall, this analysis highlights that while current privacy policies partially align with the DPDP framework, full compliance requires addressing a few key gaps, especially in multilingual accessibility, explicit consent mechanisms, and an easy consent withdrawal process.

\subsection{Ablation Study}
To establish the need for structured attribute extraction as done in \modelname, we compare it against direct prompting, where the LLM is asked directly to check DPDP compliance when provided with a privacy policy text and the DPDP act text. The prompt also asks for clear and concise reasoning. The result we present for such direct prompting primarily focuses on the two attributes that are particularly challenging due to their implicit nature: \att{easy\_consent\_withdrawal} and \att{consent\_is\_unambiguous}. These attributes require interpreting nuanced legal language and are often not explicitly stated in privacy policies.

It was observed that for the attribute \att{easy\_consent\_withdrawal}, direct prompting achieves an accuracy of $68\%$ (vs $76\%$ in \modelname), corresponding to $17$ correct predictions out of $25$ policies. While this performance improvement is moderate, it indicates that \modelname\ can identify explicit mentions of withdrawal mechanisms in several cases. However, errors arise when the policy describes withdrawal procedures indirectly them within broader legal text, making it difficult for the model to reliably assess ease of withdrawal.

For the attribute \att{consent\_is\_unambiguous}, direct prompting performs significantly worse, achieving $52\%$ (vs $64\%$ for \modelname) accuracy, meaning $13/25$ correct. This lower performance reflects the inherent difficulty of determining whether consent is expressed through clear affirmative action. In the absence of explicit attribute schema, the model often fails to distinguish between explicit and implicit consent formulations.

These results highlight the limitations of direct prompting for compliance-related attribute inference. In contrast, \modelname\ decomposes the task into well-defined base attributes and leverages rule-based reasoning for checking DPDP compliance. This structured formulation reduces ambiguity in attribute extraction and improves the reliability of downstream compliance decisions. Such a behavior is consistent with prior work showing that decomposing complex tasks into simpler, structured subproblems improves the performance of LLMs~\cite{khot2023decomposed}.



\section{Related Work}
\label{sec:related_work}

Work on privacy can be broadly divided into two categories. The first is related to anonymization during data sharing \cite{Zimmeck2019MAPSSP}, which spans research over a few decades starting from the foundational work on k-anonymity \cite{10.1145/275487.275508}, l-diversity \cite{10.1145/1217299.1217302} and t-closeness \cite{DBLP:conf/icde/LiLV07}, to more recent approaches for differential privacy \cite{Algorithmic_foundations_of_DP} and local differential privacy \cite{kasiviswanathan2011can}. The second category (to which \modelname\ belongs) involves compliance with privacy laws especially in the context of what enterprises are allowed to do with the data of their users~\cite{barth2006privacy}. 

One of the early laws governing privacy is GDPR. It has several clauses and an extensive body of work exists for addressing those. For instance, Kubicek et al. \cite{DBLP:conf/www/0001MBB24} study the problem of automating website registration for GDPR compliance. Cejas et al. \cite{DBLP:journals/tse/CejasAAB23} use NLP techniques for checking agreements related to data processing with the requirements of GDPR. Klein et al. \cite{DBLP:conf/ccs/KleinRBKJ23}, on the other hand, show a pathway for runtime checking of GDPR compliance for existing applications. LLMs have also been considered for GDPR compliance. For instance, Cory et al. \cite{DBLP:journals/popets/CoryRKRHK26} have recently used word-level annotation  with the help of LLMs for GDPR transparency compliance in privacy policies.

A recent research has explored automating privacy violation detection through the use of Contextual Integrity theory \cite{li2025privacychecklistprivacyviolation}. This theory, which views privacy not just as secrecy but as the flow of information governed by specific contextual norms, offers a structured approach to identifying potential privacy violations. The \textit{Privacy Checklist} method applies CI to assess privacy policies, providing a systematic way to detect privacy issues while emphasizing the importance of context and ethical considerations in privacy law enforcement. It laid the groundwork for incorporating CI parameters into automated compliance detection. Another closely related work is \textit{PrivaCI-Bench} framework \cite{li2025privacibenchevaluatingprivacycontextual}, which utilizes CI theory to assess privacy policies and their alignment with legal standards, offering a model to evaluate how well LLMs understand privacy contexts. It demonstrates the application of CI parameters to annotate privacy-related texts and evaluate their compliance with privacy laws. 
While the framework offers a robust foundation for embedding CI into privacy compliance systems, replicating the CI-based annotation system is not particularly straightforward. 

There is also extensive work on specific clauses of GDPR, e.g.,  the Right to be Forgotten (RTBF) \cite{DBLP:conf/ccs/CohenSSV23}\cite{DBLP:conf/ieeecai/LoboGS23}. In a recent work, Parikh et al. showed how RTBF can be enforced across collaborating domains using permissioned blockchains \cite{DBLP:conf/dbsec/ParikhSAV25}.
Besides GDPR, in the health sector, HIPAA is a well-known standard. Several approaches have been proposed targeting different aspects of HIPAA compliance. Liginlan et al. \cite{DBLP:journals/compsec/LiginlalSKF12} use Norman's action theory for compliance with HIPAA privacy rules. In contrast, Madine et al. \cite{DBLP:conf/acsac/MadineASJ25} use Zero-Knowledge Proofs for GDPR and HIPAA compliance in healthcare applications. 
On CCPA, Acquah et al. \cite{DBLP:conf/dgo/AcquahGC24} examine the effects of CCPA on data breach incidents in organizations. There are also studies combining GDPR and CCPA. For instance, Hosseini et al. \cite{DBLP:journals/popets/HosseiniUDH24} do a bilingual longitudinal analysis of privacy policy to measure impacts of booth GDPR and CCPA. 

From the above survey, it can be concluded that 
there is no attempt yet on determining if privacy policies of enterprises are compliant with the various provisions of the DPDP Act. 


\section{Conclusions and Future Work}
\label{sec:conclusion}
\modelname\ is the first ever work on showing how the provisions of the DPDP Act can be checked for compliance in any privacy policy. The ABAC framework for privacy policy checking is also a novel idea that has not been used in the context of any other privacy laws as well. Compared to standard CI formulations, the ABAC representation provides greater flexibility and compositionality, supporting richer predicates and extensible policy specifications. As a result, \modelname\ yields a more uniform and operational mechanism for verifying compliance of privacy policies with legal requirements, while naturally supporting extensions beyond compliance checking. Our results highlight two key observations. First, representing both privacy laws and policies in a unified framework like \modelname\ enables precise, attribute-level compliance analysis. Second, while LLMs are highly effective for extracting explicit compliance signals, ambiguity in natural language remains a primary source of error. Experimenting with other LLMs can give more insights into how to reduce such errors. 

We plan to extend \modelname\ to cover the remaining sections of DPDP and also establish its applicability in other privacy laws. Beyond privacy policies, we plan to explore how the framework can be meaningfully applied in resolving court cases related to DPDP violations. In view of the fact that DPDP needs consent related information to be made available in multiple languages, providing such multi-lingual support will be a natural extension of \modelname. 

\clearpage
\bibliographystyle{splncs04}
\bibliography{references.bib}

\appendix
\section{Appendix}

In this Appendix, we present the LLM prompts used in our work and an additional set of results of our experiments.

\subsection{Prompt used for attribute inference}
\label{subsec:prompt}
{\small
\textbf{Role:} \\
You are an expert in Privacy Law (specifically the Digital Personal Data Protection Act of India) and in Attribute-Based Access Control (ABAC) systems.
\textbf{Task Overview:} \\
Analyze a given privacy policy and infer the values of predefined attributes used in an Attribute-based Model.
\textbf{Input:}
\textbf{1. Privacy Policy} \\
A block of text containing the privacy policy of a company.
\textbf{2. Attribute Definitions (JSON)} \\
A JSON array where each element contains:
 - attribute\_name
 - values
 - description
Example:
\begin{verbatim}
[{
    "attribute_name": "law_applicable",
    "values": ["true", "false"],
    "description": "Whether the DPDP Act applies to this 
                    data processing activity"},
  {
    "attribute_name": "consent_status",
    "possible_values": ["not_given", "active", "withdrawn"],
    "description": "Status of consent from the data principal"
}]
\end{verbatim}
\textbf{Instructions:}
\begin{enumerate}
    \item Carefully read the privacy policy.
    \item Determine whether the policy explicitly or implicitly provides information related to each attribute.
    \item If explicitly stated, assign the corresponding value.
    \item If logically inferable, assign the inferred value.
    \item Strictly choose values only from the provided \texttt{possible\_values}.
\end{enumerate}
\textbf{Output Format:}
\begin{verbatim}
[{
    "attribute_name": "law_applicable",
    "inferred_value": "true",
    "justification": "The policy states that services are offered to users in India, 
    which brings the processing under the DPDP Act."
  },{
    "attribute_name": "consent_status",
    "inferred_value": "active",
    "justification": "The policy states that personal data is processed after 
    obtaining user consent during registration."
  }]
\end{verbatim}
\textbf{Constraints:}\\
 - Do NOT introduce new attributes.\\
 - Do NOT invent attribute values not listed in \texttt{possible\_values}.\\
\textbf{Privacy Policy:} \\
\texttt{<PRIVACY\_POLICY\_TEXT>} \\
\textbf{Attribute Definitions:} \\
\texttt{<ATTRIBUTE\_JSON>}
}\\
\emph{Note: The full prompt with attribute definitions can be found in the compliance model backend code linked in the main paper.}

\subsection{Full Per-Policy and Per-Attribute Performance}
\label{subsec:per_policy_performance}

These are listed in the two tables.

\begin{table}[t]
\centering
\caption{Performance metrics per privacy policy.}
\scriptsize
\begin{tabular}{l c c c c}
\toprule
Policy & Acc & Prec & Rec & (TP, TN, FP, FN) \\
\midrule
A1 & 0.9375 & 1.0000 & 0.9167 & (11,4,0,1) \\
A2 & 0.9375 & 1.0000 & 0.9091 & (10,5,0,1) \\
B1 & 1.0000 & 1.0000 & 1.0000 & (11,5,0,0) \\
B2 & 0.9375 & 1.0000 & 0.9167 & (11,4,0,1) \\
B3 & 0.9375 & 1.0000 & 0.9231 & (12,3,0,1) \\
C1 & 1.0000 & 1.0000 & 1.0000 & (11,5,0,0) \\
C2 & 1.0000 & 1.0000 & 1.0000 & (11,5,0,0) \\
F1 & 1.0000 & 1.0000 & 1.0000 & (11,5,0,0)\\ F2 & 0.9375 & 1.0000 & 0.9091 & (10,5,0,1) \\
G1 & 0.9375 & 1.0000 & 0.9167 & (11,4,0,1) \\
H1 & 0.9375 & 0.9231 & 1.0000 & (12,3,1,0) \\
H2 & 0.9375 & 1.0000 & 0.9167 & (11,4,0,1) \\
I1 & 0.9375 & 0.8889 & 1.0000 & (8,7,1,0) \\
L1 & 0.8750 & 0.9167 & 0.9167 & (11,3,1,1) \\
N1 & 0.8750 & 0.8462 & 1.0000 & (11,3,2,0) \\
N2 & 0.9375 & 0.9091 & 1.0000 & (10,5,1,0) \\
N3 & 1.0000 & 1.0000 & 1.0000 & (12,4,0,0) \\
P1 & 1.0000 & 1.0000 & 1.0000 & (11,5,0,0) \\
P2 & 0.9375 & 1.0000 & 0.9231 & (12,3,0,1) \\
P3 & 1.0000 & 1.0000 & 1.0000 & (11,5,0,0) \\
S1 & 0.9375 & 0.9167 & 1.0000 & (11,4,1,0) \\
U1 & 1.0000 & 1.0000 & 1.0000 & (13,3,0,0) \\
W1 & 0.8750 & 1.0000 & 0.8333 & (10,4,0,2) \\
X1 & 0.9375 & 0.9167 & 1.0000 & (11,4,1,0) \\
Z1 & 0.8125 & 0.7500 & 1.0000 & (9,4,3,0) \\
\bottomrule
\end{tabular}
\label{tab:compoldetail}
\end{table}


\begin{table}[t]
\centering
\caption{Performance metrics per attribute.}
\scriptsize
\begin{tabular}{l c c c c}
\toprule
Attribute & Acc & Prec & Rec & (TP, TN, FP, FN) \\
\midrule
consent\_is\_freely\_given & 0.9600 & 0.9600 & 1.0000 & (24,0,1,0) \\
consent\_is\_informed & 1.0000 & 1.0000 & 1.0000 & (25,0,0,0) \\
consent\_is\_specific\_to\_purpose & 1.0000 & 1.0000 & 1.0000 & (25,0,0,0) \\
consent\_is\_unambiguous & 0.6400 & 0.2500 & 0.1429 & (1,15,3,6) \\

consent\_notice\_how\_to\_complaint\_to\_board & 0.9600 & 0.6667 & 1.0000 & (2,22,1,0) \\
consent\_notice\_how\_to\_exercise\_rights\_sec13 & 0.9200 & 1.0000 & 0.9130 & (21,2,0,2) \\
consent\_notice\_how\_to\_exercise\_rights\_sec6.4 & 0.8800 & 0.8800 & 1.0000 & (22,0,3,0) \\
consent\_notice\_information\_about\_\\personal\_data & 1.0000 & 1.0000 & 1.0000 & (25,0,0,0) \\
consent\_notice\_purpose\_of\_processing & 1.0000 & 1.0000 & 1.0000 & (25,0,0,0) \\
consent\_request\_contains\_contact\_details\_of\_\\dpo\_or\_equivalent & 1.0000 & 1.0000 & 1.0000 & (23,2,0,0) \\
consent\_request\_language\_all\_eighth\_schedule & 1.000 & 0.000 & 0.000 & (0,25,0,0)\\
easy\_consent\_withdrawal & 0.7600 & 0.5714 & 0.5714 & (4,15,3,3) \\
lawful\_purpose & 1.0000 & 1.0000 & 1.0000 & (25,0,0,0) \\
notice\_languages\_all\_eighth\_schedule & 1.0000 & 0.0000 & 0.0000 & (0,25,0,0) \\
offering\_service\_to\_data\_principal\_\\within\_india & 1.0000 & 1.0000 & 1.0000 & (25,0,0,0) \\
purpose\_of\_processing & 1.0000 & 1.0000 & 1.0000 & (25,0,0,0) \\
\bottomrule
\end{tabular}
\label{tab:perattribdetail}
\end{table}

\end{document}